# THE NATURE OF MICROJANSKY RADIO SOURCES AND IMPLICATIONS FOR THE DESIGN OF THE NEXT GENERATION VERY SENSITIVE RADIO TELESCOPES

K. I. KELLERMANN
National Radio Astronomy Observatory
520 Edgemont Road
Charlottesville, VA 22903
E-mail:  kkellerm@nrao.edu

E. A. RICHARDS
NRAO and University of Virginia, Charlottesville, VA
*Present address:* ASU, Dept. of Physics and Astronomy, Tempe AZ
E-mail:  erichard@nrao.edu

Deep radio surveys show a population of very red counterparts of microjansky radio sources, which are unidentified to I = 25 in ground based images and I = 28.5 in the Hubble Deep Field.  This population of optically faint radio sources, which comprises about 20 percent of the microjansky samples, may be dust enshrouded starburst galaxies, extreme redshift or dust-reddened AGN, or due to displaced radio lobes.  Even deeper radio surveys, which will be made possible by next generation radio telescopes such as the Expanded VLA or the Square Kilometer Array, will reach to nanojansky levels which may be dominated by this new population, but only if special care is taken to achieve high angular resolution and dynamic range better than 60 db.  This will require array dimensions up to 1000 km to achieve confusion limited performance at 1.4 GHz and up to 10,000 km at 300 MHZ.  But, even then, the ability to study individual nanojansky radio sources may be limited by the finite extent of the sources and the consequential blending of their images.

## 1.  Introduction

Over the past decade we have been using the VLA for deep radio surveys in order to better understand the nature of microjansky radio sources and, in particular, their implication for star formation at early epochs (Fomalont et al 1988; Fomalont et al. 1991).  Kellermann et al. (in preparation) and Windhorst et al. (1995) discuss a deep VLA survey at 8.4 GHz centered on SSA 13, which detected 39 radio sources above a completeness level of 7 µJy.  For the past few years we have concentrated on the Hubble Deep Field where our sensitive radio observations at 8.5 GHz (Richards et al. 1998) and 1.4 GHz (Richards 1999a, Muxlow et al. 1999) complement the deep HST imaging and sensitive ground based spectroscopy.  In addition, observations with NICMOS (Thompson et al. 1999), ISO (Aussel et al. 1999), and SCUBA (Hughes et al 1999) provide data in the near IR, mid IR, and FIR (sub-mm) wavelengths respectively.  The observations at radio wavelengths have sufficient angular resolution to resolve blends, which allow unambiguous identification with faint objects seen in other wavelength bands.

The VLA surveys of the HDF have been made at two frequencies, 1.4 and 8.5 GHz, with angular resolution between 2 and 6 arcseconds as summarized in Table 1.  The 1.4 GHz VLA observations have been supplemented with MERLIN observations, which improve the resolution to 0.2 arcseconds (Muxlow et al., 1999 and in preparation).  The combined VLA+MERLIN image has an rms noise level of about 4 µJy and is the most sensitive image ever made at this frequency.

Table 1
VLA HDF Surveys

| λ(cm) | τ(hours) | σ(μJy) | $S_{lim}$(μJy) | N | N (HDF) |
|---|---|---|---|---|---|
| 3.6 | 192 | 1.6 | 8 | 40 | 7 |
| 20 | 50 | 7.5 | 40 | 371 | 7 |

## 2. Identification Statistics

There are a total of 79 sources located in regions covered by good space or ground based optical material. Of these, 11 are in the HDF above either the 3.6 or 20 cm completeness limit shown in Table 1. Figure 1 shows the distribution of optical magnitude for these 79 sources. For the identified sources, the optical counterparts are mostly galaxies with active star forming regions at moderate redshift or weak AGN associated with massive central engines. This classification is made on the basis of the radio spectrum, the angular size and morphology of the radio emission, as well as optical, IR, and submillimeter imaging, color, spectra, and morphology, but it is not always unambiguous.

For the most part, the millijansky radiation is due to synchrotron emission driven by a massive central engine in the nuclei of an elliptical galaxy, while the microjansky sources are found mostly in exponential, distorted, interacting, or merging systems suggestive of synchrotron radiation driven by starbursts and subsequent coalesced supernovae activity. There are no powerful FR II radio galaxies among the microjansky radio sources and only very few FR I radio galaxies.

As shown in Figure 1, 11 radio sources remain unidentified down to the optical limit of $I = 25.5$ in the flanking fields, while three of the 11 sources which lie within the HDF remain unidentified down to the HDF limit of $I = 28.5$.

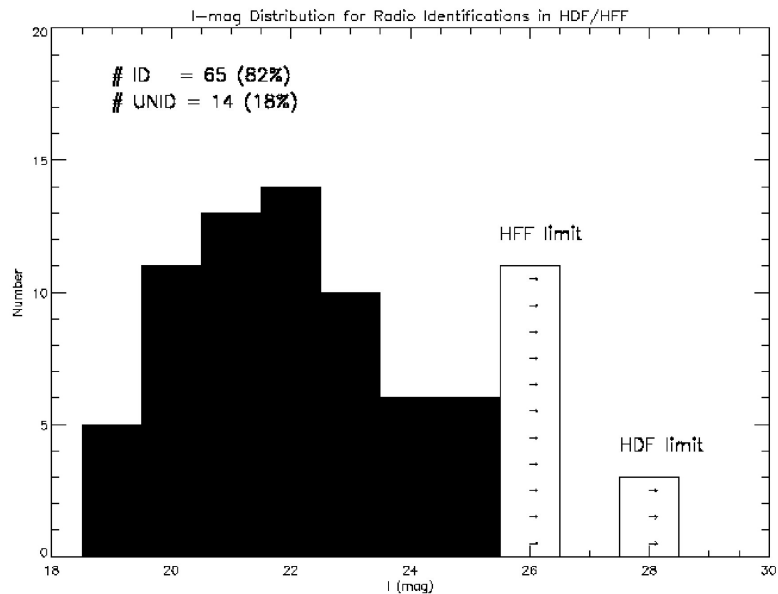

*Figure 1. Histogram showing the I magnitude distribution of radio sources in the HDF region*

There has been considerable recent attention to the reported submmillimeter SCUBA observations in the HDF (Hughes et al. 1998). The apparent resolution of the FIR/submmillimeter background into discrete sources (Smail et al. 1997, Hughes et al. 1998) is particularly exciting. However, due to the poor angular resolution of the SCUBA observations, source positions are poorly determined; a situation which is exacerbated by confusion. Even in the HDF, despite the wealth of supplementary data, the identification of the submmillimeter sources remain ambiguous. Nevertheless, it is of interest that two VLA radio sources, which are located close to SCUBA sources, are unidentified in the HDF images.

## 3. Optically Faint Counterparts

What is the nature of the unidentified radio sources? There are several possibilities.

- They are FRI or FRII radio galaxies at moderate redshift. But, this would mean that M > -14 (for z=1) whereas typical FR I or FR II radio galaxies have M ~ -20.
- The radio source may be a displaced lobe of an asymmetric source. However other microjanksy radio sources consist of only a single component coincident with the optical counterpart.
- The optical counterpart may be at such a high redshift, z > 7, that it is an I band dropout.
- The optical counterpart may be obscured by dust.
- The unidentified sources may represent a new population of objects.

## 4. Comments on Specific Sources

We discuss five radio sources with particularly interesting optical/IR counterparts.

*VLA 123646+621226* can be characterized as follows: $S_{20}$ = 72 µJy, $S_{8.5}$ = 8.5 µJy, $\alpha$ = 0.9. It is well resolved by the VLA beam with an angular size, $\theta$ = 2.9 arcseconds. No optical counterpart is visible down to the HDF limit of I=28.5. Deep near IR imaging by Thompson et al. (1999) place limits of J > 29 and H > 29. Positions measured independently from the 1.4 and 8.5 GHz images agree to within 0.2 arcseconds consistent with their standard errors, so that there is little doubt as to the reality of this diffuse steep spectrum radio source. The 1.4 GHz position is RA(J2000) = 12:36:46.698±027, Dec(J2000) = +62:12:26.54±0.19.

*VLA 123656+621207* is an unidentified radio source near the edge of the HDF with I > 28.5. Like VLA 123646+621226, it has a steep radio spectrum with $\alpha$ > 1.3. and $S_{1.4}$ = 46 µJy, but it is not detected at 8.5 GHz with an upper limit of 5 µJy. IR NICMOS imaging show no counterpart to limits of J > 25 and H > 25 (Dickinson, private communication) or K > 24 (Barger et al. 1999). VLA 123656+621207 is located 3 arcseconds from the second brightest SCUBA source HDF 850.2 which is 3.8 mJy at 850 microns (Hughes et al. 1998). Considering the uncertainty in the SCUBA position, the 850 micron SCUBA source is probably the same as the radio source VLA 123656+621207, although the radio position excludes the optical identification proposed by Hughes et al.

*VLA 123642+621331* is located in one of the HDF flanking fields. It is a relatively strong source, with $S_{1.4}$ = 430 µJy, $S_{8.5}$ = 80 µJy, and $\alpha$ = 0.9. Figure 3 shows radio contours constructed from the combined MERLIN-VLA data overlayed on the HST flanking field image. No optical counterpart is visible to I = 25.5. A CFHT K band image shows a surprisingly bright, K = 20.8 galaxy, so that I-K > 4.7 (Barger et al 1999). NICMOS observations confirm the existence of a very red counterpart, with J=23.8 and H=22.1 and a disk profile, suggesting a dust enshrouded nuclear starburst (Waddington et al., in preparation). It is not detected at 850 microns to a limit of 2 mJy (Hughes et al. 1998). Most of the radio flux density is confined to a compact region < 0.1 arcseconds in diameter.

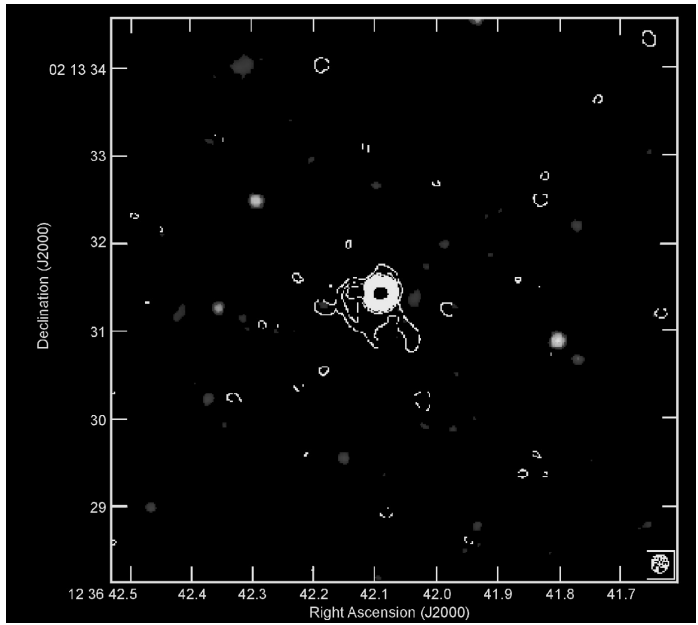

*Figure 2. Radio contours of VLA 123642+621331 from the VLA-MERLIN image overlayed on the HFF optical image.*

*VLA 123651+621226:* The brightest 850 micron SCUBA source, HDF850.1 (RA=12:36:52.32±0.10, Dec = +62:12:26.3±0.7, $S_{850}$= 7 mJy) is located close to the weak radio source VLA 123651+621226 (RA=12:36:51.960±0.061; Dec = +62:12:26.09±0.42, $S_{1.4}$ < 23 µJy, $S_{8.5}$ = 8.6 µJy, and α < 0.5). Downes, et al. observe a 2.2 ± 0.3 mJy source at 1.3 mm at RA = 12:36:51.98±0.07, Dec = 62:12:25.7±0.3). Downes et al. suggest an identification either with HDF3-593.0 which may be a lensed arc of a z = 1.7 galaxy by the nearby very red elliptical galaxy, HDF 3-586.0 (R-K=5) or with HDF3-586.0 itself. More likely, however, HDF850.1 is a blank field object at very high redshift (z > 3), as suggested by the radio and submmillimeter measurements.

*VLA 123651+621221:* The nearby $S_{1.4}$ = 63 µJy, $S_{8.5}$ = 17 µJy, α = 0.7 radio source VLA 123651+621221 was originally identified with the bright (I = 22) spiral galaxy on the basis of the low resolution VLA 8.5 GHz image. (Richards et al. 1998). However the combined VLA-MERLIN high resolution 1.4 GHz image suggests an identification with a nearby very faint feature near the HDF limit. NICMOS observations by Dickinson et al. (private communication) show this feature to be very red (J-H=1.9). Downes et al. see a 3 sigma 1.3 mm source ($S_{1.3}$=1.2 mJy) within 1 arcseconds of VLA 123651+621221. VLA 123651+621221 may be a highly redshifted dust enshrouded system. Most likely the bright SCUBA source HDF850.1 is a blend of VLA 123651+621226 and VLA 123651+621221.

**5. Summary of HDF Radio Observations**

Sensitive high resolution radio imaging have disclosed a population of optically faint very red counterparts which may be unusually strong at submmillimeter wavelengths. However, the radio observations are essential to locate optical and IR counterparts, as the low resolution observations made with SCUBA or other bolometer arrays have inadequate angular resolution and are subject to confusion. The VLA radio observations combined with the submillimeter SCUBA observations suggest redshifts for the submillimeter population in the range 1<z <3 (Barger, Cowie, & Richards 1999).

Even using the current VLA, we are able to detect weak radio sources whose optical counterparts are apparently beyond the limits of the largest optical and infra-red telescopes. More sensitive observations in the IR and submmillimeter bands will be important to understand the nature of these optically faint radio sources. Meanwhile, following the planned upgrade of the VLA, radio sources weaker by a factor of 5 to 10 will be observed. The SKA will have the sensitivity to reach sources as weak as 100 nJy in a single track and perhaps 25 nJy in long integrations.

**6. Implications for the Design of the SKA and Other Next Generation Radio Telescopes**

Increased collecting area, alone, is not adequate to reach the lowest levels of sensitivity. The SKA will also need to have low noise broadband receivers to complement the large collecting area. But, to fully exploit the potential sensitivity of the SKA, it will be necessary to also have high angular resolution to minimize primary beam confusion and extremely low sidelobe levels to avoid spurious responses from "strong" sources, which may be far removed from the target position.

For example, we note that the Arecibo radio telescope has about 2.5 times the collecting area of the VLA at 20 cm. But, the weakest sources that can be reliably observed at Arecibo is limited by confusion to about 10 mJy at 20 cm, whereas the VLA with its smaller collecting area but second of arc resolution can reliably observe sources one thousand times weaker. The 20 cm VLA NVSS (Condon et al. 1998) was made in the D-configuration, which has overall dimensions about one km. The observed rms confusion in the NVSS is about 100 µJy (Condon, private communication) so if the SKA were literally configured into a square km, the weakest sources, which could be reliably observed at 20 cm, would be about 1 mJy. SKA specifications call for an rms noise level of about 20 nJy in 12 hours. Deep integrations might run for a few hundred hours and would be about a factor of four more sensitive. So the noise limited detection levels will be about 100 nJy and 25 nJy in 12 hours and 200 hours respectively.

Source counts at 20 cm indicate a source density at 1 µJy of about 100 sources per sq arcmin. Extrapolation gives 2,500 and 60,000 sources per sq arcmin stronger than 100 nJy and 25 nJy respectively. This corresponds to a mean separation between sources of 6, 1 and 0.5 arcseconds at 1000, 100, and 25 nJy respectively. Using a guideline of at least 25 beam areas per sources, means that a resolution of 1, 0.25, and 0.1 arcseconds respectively is required just to separate individual sources. Even with this resolution, the ability of the SKA to adequately separate individual sources will depend critically on the nature of the nanojansky sources themselves. If the nanojansky sources are unresolved "point" sources, then the above arguments are relevant. If, however, they have galactic dimensions of 10 kpc or dimensions of even a few hundred parsecs, characteristic of active starforming regions, then typical angular sizes at high redshift will be ~ 1 arcseconds and ~ 0.05 arcseconds respectively in any realistic world model. Then, no matter what the instrumental angular resolution, the ability to reach weak sources will be limited by this natural confusion where sources begin to overlap in the sky. Assuming a limiting source density corresponding to 25 characteristic source sizes, we estimate that the minimum flux density will be ~ 1 µJy if the characteristic source size is 1 arcseconds (10 kpc) and 10 nJy if it is 0.05 arcseconds (500 parsecs). These arguments imply array dimensions of 100, 400, 1000 km at 1.4 GHz to reach point source levels of 1000, 100, and 25 nJy respectively. Thus, intrinsic dimensions up to a few hundred parsecs will not degrade the performance of the SKA above 10 nJy at 1.4 GHz provided that the array is at least 1000 km in extent.

From the VLA observations with 2 arcseconds resolution, Richards (1999) finds a 1.4 GHz mean angular size of 1.8 arcseconds for microjansky sources, characteristic of galactic disks at moderate redshift. The combined VLA-MERLIN observations, which have 0.2 arcseconds resolution, suggest a characteristic size about 1.5 arcseconds (Muxlow et al. 1999). If this is characteristic of the nanojansky population, then the weakest observable sources might be as strong as a few hundred nanojansky at 1.4 GHz and about 1 µJy at 300 MHz.

The number density of sources scales with wavelength as $(\lambda_2/\lambda_1)^{\alpha k}$ where α is the spectral index and x is the exponent of the integral number-flux density relation. Both x and α depend on flux density and frequency. Using typical values of x = 1.4 (Richards 1999) and α = 0.7, αx ~ 1, and so the source density is about 500 per sq arc min at 300 MHz for S > 1 µJy, and about $10^6$ sources per sq arcmin for S > 25 nJy. Thus, to avoid confusion limited performance at 300 MHz a resolution of at least 0.5 arcseconds for S > 1 µJy, and 0.05 arcseconds for S > 25 nJy will be required. These numbers imply array dimensions of 1000 and 10,000 km respectively.

These estimates depend on an extrapolation of current microjansky source counts by more than two orders of magnitude. Some constraints from the FIR background (Haarsma & Partridge 1998) and the CMB (Windhorst et al. 1993) suggest that the counts may begin to converge below about 1 µJy. If so, the actual source density will be smaller than described above, or they may be larger if new populations appear. Deep surveys with the upgraded VLA will shed some light on the sub-microjansky source counts as well as on their angular size distribution.

Adequate suppression of sidelobes will present an additional challenge. The level of required suppression will depend on the primary beam field-of-view. The strongest source present within the field-of-view will be when N (S) Ω = 1. Taking Ω = one sq degree at 20 cm, the strongest source within the field of view may be about 100 mJy at 1.4 GHz. We want the spurious response to be less than the thermal noise. This puts a requirement on the dynamic range of 60-70 db to reliably reach sources as weak as 25 nJy. At 300 MHz the strongest source within a 1 degree field will be about 5 times stronger, so a dynamic range close to 70 db will be needed to avoid contamination from spurious responses on the deepest surveys. If the field of view scales with wavelength, then at 300 MHz it will be about 25 square degrees and an additional 10 db suppression will be needed across this field.

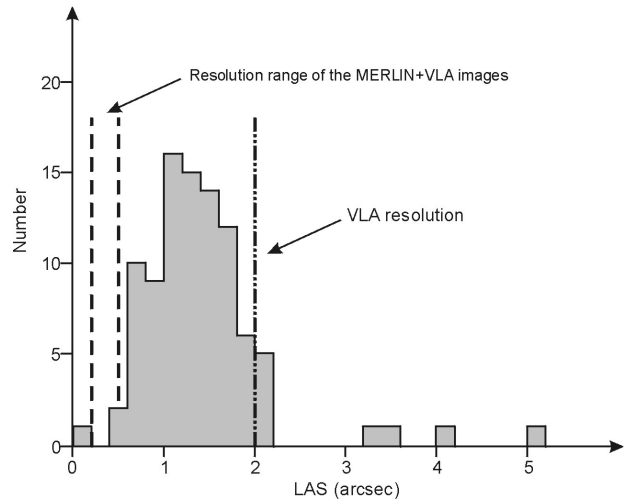

*Figure 3. Histogram showing the distribution of angular sizes for radio sources detected in the combined VLA/MERLIN survey of the HDF region (from Muxlow et al. 1999).*

This will be a formidable challenge for the design of the SKA. The best dynamic range, which has been achieved to date, is in the range of 40 to 50 db for an isolated source located near the phase center of the array. Full beam imaging at the VLA appears to be limited to a dynamic range of a few thousand using current techniques. Improvement in dynamic range for full beam imaging will require careful attention to pointing stability, closure errors, imaging algorithms, and the rejection of interference.

The strawman SKA design consists of a central "element" containing 80 percent of the collecting area some 50 km in diameter plus nine outrigger "elements," three on each of three arms extending out to 500 km. The outer nine elements each contain about 2 percent of the total collecting area, or about that of the VLA. This type of configuration can give good resolution corresponding to the 500 km dimensions, but only if the "ten" elements are uniformly weighted, so that the sensitivity of such an image would be reduced by a factor of about 5. The dynamic range and image quality of this "ten" element array will also be limited and cannot meet the requirements discussed above. To obtain the full sensitivity of the SKA would require natural weighting, in which case the array would be dominated by the central element and the resolution would correspond to the 50 km dimensions of the central element, or about equivalent to that of the VLA in its A-configuration.

The VLA and the WST are able to cover a wide range of resolutions and surface brightness sensitivity by moving the elements among different configurations. This probably will not be realistic with the SKA. Extensive study will be needed to find a configuration for the SKA that optimizes any compromises between sensitivity, resolution, and surface brightness sensitivity, exploiting where feasible, multifrequency synthesis techniques to enhance the u,v coverage. More extensive observations of sub-microjansky sources with the Expanded VLA will needed to better determine their sky density, luminosity function, and angular size distribution in order to see if it will be feasible to reach the 25 to 100 nJy sources with the SKA.

**Acknowledgements**

We thank our colleagues E. Fomalont, B. Partridge, R. Windhorst, I. Waddington, and T. Muxlow who have collaborated in this program, as well as A. Barger, L. Cowie, and M. Dickinson who have kindly allowed us to use their data in advance of publication.

The National Radio Astronomy Observatory is a facility of the National Science Foundation operated under cooperative agreement by Associated Universities, Inc.